\documentclass[aps,twocolumn,prb]{revtex4-2}

\usepackage{amsmath}
\usepackage{listings}
\usepackage{graphicx}
\usepackage{dcolumn}
\usepackage{bm}
\usepackage{graphics}
\usepackage{bbold}
\usepackage{epsfig,color}

\usepackage{mathtools,amssymb}
\usepackage{braket}
\usepackage{stmaryrd}
\usepackage{bbm}
\usepackage{dsfont}

\usepackage[breaklinks,colorlinks,bookmarks=false,citecolor=red,linkcolor=blue,urlcolor=magenta]{hyperref}

\begin{document}

\title{Multilevel Quantum Dynamics: A Parafermionic Truncated Wigner Approach}

\author{Javad Vahedi}
\email[]{javahedi@gmail.com}
\affiliation{Institute of Condensed Matter Theory and Optics, Friedrich-Schiller-University Jena, Max-Wien-Platz 1, 07743 Jena, Germany}

\author{Martin Gärttner}
\email[]{martin.gaerttner@uni-jena.de }
\affiliation{Institute of Condensed Matter Theory and Optics, Friedrich-Schiller-University Jena, Max-Wien-Platz 1, 07743 Jena, Germany}

\begin{abstract}
We introduce the parafermionic truncated Wigner approximation ($p$TWA),
a semiclassical phase-space method for studying the nonequilibrium
dynamics of multilevel quantum systems, including $\mathbb{Z}_n$ Fock
parafermions. Motivated by the fractional exchange statistics of Fock
parafermions, we express the Hamiltonian in a basis of local Hubbard
operators that closes under commutation. This representation yields a
Lie--Poisson formulation in which stochastic sampling of the initial
state is followed by deterministic semiclassical evolution.

We benchmark the method for single-site clock dynamics, a fully
connected $\mathbb{Z}_n$ clock model that generalizes the
Lipkin--Meshkov--Glick model, long-range $\mathbb{Z}_3$ clock chains,
and clean and disordered one-dimensional Fock parafermion chains.
The method captures qualitative features of the quantum benchmarks,
including excitation spreading, disorder-induced suppression of
transport, and long-time imbalance plateaus. In the single-hopping
parafermion chains, it predicts slower domain-wall relaxation than
time-evolving block decimation.
\end{abstract}

\maketitle

\section{Introduction}
Understanding nonequilibrium dynamics in interacting quantum many-body systems is a central challenge across quantum science, with implications ranging from quantum information processing to topological matter and emergent statistical phenomena~\cite{Polkovnikov2011,Calabrese2016,DAlessio2016}.
The exponential growth of the Hilbert space with system size severely limits exact numerical simulations, motivating the development of approximate yet scalable theoretical frameworks. Tensor-network methods~\cite{Schollwck2011,Paeckel2019} and quantum Monte Carlo approaches~\cite{Ceperley2003,Profumo2015} have achieved remarkable success in specific regimes, but face intrinsic challenges for real-time dynamics, long-range interactions, and systems with unconventional exchange statistics~\cite{Eisert2015}.

Semiclassical phase-space methods provide a complementary route for studying large quantum systems. Among these, the truncated Wigner approximation (TWA) maps quantum operators to classical phase-space variables whose stochastic evolution captures leading quantum fluctuations~\cite{Polkovnikov2010}. Originally developed for bosonic and spin systems, TWA has enabled studies of coherence dynamics, thermalization, and correlation spreading in systems beyond the reach of exact techniques~\cite{Blakie2008,Wurtz2018}.

TWA-type phase-space methods have also been generalized to finite-dimensional multilevel systems. In particular, generalized discrete truncated Wigner approaches formulate the dynamics in terms of a complete local operator basis, such as generalized Gell-Mann matrices, and represent the corresponding quantum fluctuations through stochastic sampling of classical phase-space variables~\cite{Zhu2019}. Such formulations provide a natural semiclassical description of local $n$-level degrees of freedom and are closely related to the Hubbard-operator representation employed below.

Related algebraic constructions have enabled extensions of semiclassical
phase-space methods to fermionic systems. There, the appearance of
Grassmann-valued phase-space variables obstructs direct stochastic sampling in terms of the elementary fermionic fields. This difficulty can be circumvented by reformulating the dynamics in terms of fermionic bilinears that form closed Lie algebras---such as $\mathfrak{so}(2N)$ or $\mathfrak{so}(4N)$ in Majorana representations---allowing semiclassical evolution with ordinary probability  distributions~\cite{Davidson2017,Vahedi2025}. Together with multilevel formulations, these developments highlight the usefulness of Lie--Poisson phase spaces for semiclassical descriptions based on noncanonical local operator algebras~\cite{Kirillov2004,Littlejohn1986}.

In this work we develop such a phase-space description for $\mathbb{Z}_n$
Fock parafermions~\cite{Cobanera2014,Rossini2019,Mahyaeh2020}. Parafermions interpolate between fermions ($n=2$) and more exotic fractionalized excitations ($n>2$), and appear in a variety of contexts including clock models, topological phases, and non-Abelian defects~\cite{Wilczek1982,Nayak2008,Fendley2012,Alicea2016}. A defining feature of parafermionic models is the presence of nonlocal Jordan--Wigner strings implementing fractional exchange statistics. As a consequence, even Hamiltonians that are quadratic in parafermionic creation and annihilation operators can encode nontrivial many-body dynamics when expressed in terms of local commuting degrees of freedom.

Our construction uses the local Hubbard operators associated with the
$n$-dimensional Fock space as phase-space coordinates. At the level of the
local multilevel dynamics, this representation is related by a linear change of operator basis to generalized Gell-Mann formulations such as
GDTWA~\cite{Zhu2019}. The Hubbard representation is particularly convenient for Fock parafermions because local occupation-changing operators are expressed directly in this basis, while the fractional exchange statistics can be incorporated through the corresponding string structure. We derive the resulting semiclassical equations of motion and develop Gaussian and discrete prescriptions for sampling the initial quantum fluctuations. Below we make the relation to existing multilevel phase-space formulations explicit and then apply the construction to parafermionic many-body dynamics.

We benchmark the $p$TWA framework in a series of representative
settings. As controlled tests we first examine single-site
$\mathbb{Z}_n$ clock dynamics and the fully connected
$\mathbb{Z}_n$ clock model, where semiclassical dynamics becomes
exact in appropriate limits.
We then apply the method to interacting one-dimensional systems,
including long-range $\mathbb{Z}_3$ clock chains and disordered
$\mathbb{Z}_3$ Fock parafermion models, where we demonstrate that
$p$TWA captures the qualitative features of excitation transport,
disorder-induced suppression of dynamics, and long-time imbalance
plateaus. Finally, simulations of $\mathbb{Z}_n$ Fock parafermion chains with varying local dimension show systematic improvement of the
semiclassical description as the system approaches the classical
limit. The remainder of this paper is organized as follows.
In Sec.~\ref{sec:II} we formulate the $p$TWA framework and discuss the
construction of the associated Lie--Poisson phase space.
Section~\ref{sec:III} presents benchmark tests of the method,
while Sec.~\ref{sec:IV} applies the approach to one-dimensional
parafermionic lattice models.
We conclude in Sec.~\ref{sec:V} with a discussion of the implications and
possible extensions of the method.

\section{Truncated Wigner Approximation: From Bosons and Fermions to Parafermions}
\label{sec:II}

The Truncated Wigner Approximation (TWA) provides a semiclassical framework for simulating the nonequilibrium dynamics of quantum many-body systems. The central idea is to represent the quantum density operator $\hat{\rho}$ via its Wigner--Weyl transform $W(\mathbf{x},\mathbf{p})$ in a suitable phase space, and to approximate the quantum Liouville equation by a classical Liouville evolution of phase-space trajectories~\cite{Polkovnikov2010}. Quantum fluctuations enter through stochastic sampling of initial conditions from $W$, while the subsequent time evolution follows the classical equations of motion generated by the Weyl symbol of the Hamiltonian. This construction is exact for quadratic Hamiltonians and becomes a controlled approximation for weak nonlinearities or sufficiently short times.

\subsection{Bosonic phase space and Wigner sampling}
For bosonic systems with canonical operators $\hat{a}_i$, $\hat{a}_i^\dagger$ satisfying $[\hat{a}_i,\hat{a}_j^\dagger]=\delta_{ij}$, the Wigner function constitutes a quasi-probability distribution over the continuous phase space of complex amplitudes $\alpha_i$. In truncated Wigner approximation, expectation values of symmetrically ordered observables are obtained as phase-space averages:
\begin{equation}
\langle \hat{O}(t) \rangle = \int d^{2N}\!\alpha\, W(\alpha,\alpha^*; 0)\, O_W(\alpha(t),\alpha^*(t)),
\end{equation}
where $O_W$ denotes the Weyl symbol of $\hat{O}$ and $\alpha(t)$ evolves under classical equations of motion. The Poisson bracket structure,
\begin{equation}
\{ \alpha_i, \alpha_j^* \} = i\delta_{ij},
\end{equation}
yields Hamilton's equations,
\begin{equation}
i\hbar \dot{\alpha}_i = \frac{\partial H_W}{\partial \alpha_i^*}, 
\qquad
-i\hbar \dot{\alpha}_i^* = \frac{\partial H_W}{\partial \alpha_i},
\end{equation}
with $H_W$ the Weyl symbol of the Hamiltonian. Initial sampling from distributions such as Gaussians corresponding to coherent or squeezed states allows access to quantum correlations beyond mean-field theory. Bosonic TWA has found widespread application in ultracold-atom physics, nonlinear optics, and quantum field theory~\cite{Blakie2008,Hillery1984}.

\subsection{Fermionic Wigner formalism ($f$TWA)}

Extending the phase-space picture to fermions encounters a fundamental difficulty: the canonical fermionic operators $\hat{c}_i$, $\hat{c}_i^\dagger$ satisfy anticommutation relations $\{\hat{c}_i,\hat{c}_j^\dagger\}=\delta_{ij}$, which implies that the natural phase-space coordinates are Grassmann numbers $\xi_i$, $\xi_i^*$ obeying $\{\xi_i,\xi_j\}=0$. Grassmann-valued fields cannot be sampled using ordinary probability distributions, obstructing direct numerical implementation.

Several approaches circumvent this “Grassmann obstruction” by reformulating the fermionic dynamics in terms of bilinear operators that close under commutation. One approach employs the single-particle density-matrix representation $\rho_{ij}=\langle \hat{c}_j^\dagger \hat{c}_i \rangle$, which evolves under an effective mean-field Hamiltonian $h[\rho]$ with additional semiclassical corrections~\cite{Davidson2017,Sajna2020,Osterkorn2024}. Another formulation uses the antisymmetric Majorana correlation matrix $M_{ab}$, defined from $\hat{\gamma}_a=\hat{c}_a+\hat{c}_a^\dagger$, yielding an $\mathfrak{so}(4N)$ Lie--Poisson structure that captures fermionic semiclassics without explicit Grassmann variables~\cite{Vahedi2025}. Both approaches effectively replace the Grassmann phase space with a manifold of bilinear expectation values, enabling tractable semiclassical simulations of interacting fermionic systems.

\subsection{Parafermionic Wigner formalism ($p$TWA)}
We now extend the semiclassical Wigner framework to $\mathbb{Z}_n$ \emph{Fock parafermions} (FPFs), which generalize fermions ($n=2$) to fractionalized excitations with nontrivial exchange statistics for $n>2$. Local FPF creation and annihilation operators $f_j^\dagger$ and $f_j$ obey the nilpotency condition
\begin{equation}
f_j^n = (f_j^\dagger)^n = 0,
\end{equation}
and acquire fractional exchange phases
\begin{equation}
f_j f_k = \omega\, f_k f_j , \qquad (j<k),
\qquad \omega = e^{2\pi i/n}.
\end{equation}
On a single site, Fock parafermions satisfy the generalized on-site constraints
\begin{equation}
f_j^{\dagger m} f_j^{m} + f_j^{\,n-m} f_j^{\dagger(n-m)} = \mathbf{1},
\qquad m=1,\dots,n-1,
\label{eq:FPF_onsite}
\end{equation}
which reduce to the canonical fermionic anticommutation relation for $n=2$, but define a genuinely nonlinear on-site algebra for $n>2$. As a consequence of the nonlinear on-site algebra, the operators  $f_j$ and $f_j^\dagger$ do not generate a closed finite-dimensional Lie algebra under commutation. Commutators produce higher-order operator monomials, preventing a linear Poisson-bracket correspondence and rendering a direct semiclassical phase-space formulation in terms of parafermionic operators intractable.

\begin{figure*}[t]
\centering
\includegraphics[width=\linewidth]{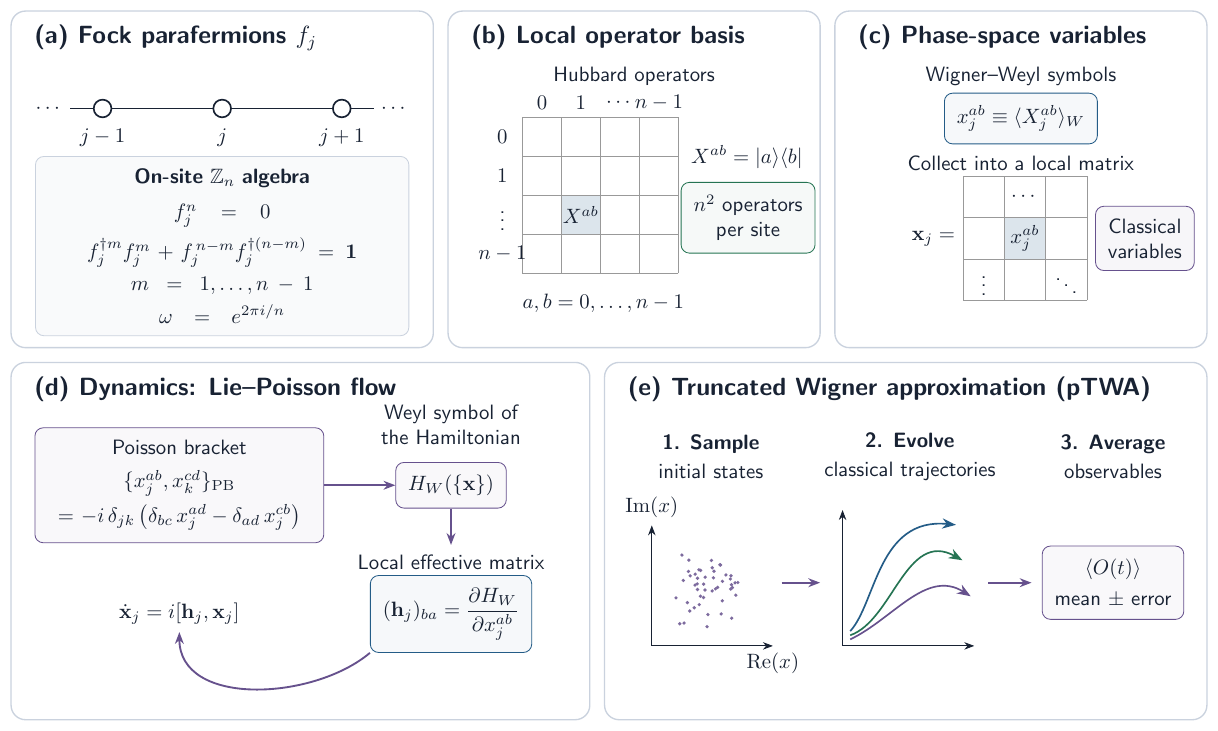}
    \caption{
    Schematic overview of the parafermionic truncated Wigner approximation
    ($p$TWA). 
    (a) Fock parafermions with a finite local Hilbert space and nontrivial
    $\mathbb{Z}_n$ algebra.
    (b) The local operator space is represented in the complete Hubbard basis
    $X_j^{ab}=|a\rangle\langle b|$.
    (c) Hubbard operators are mapped to classical Wigner--Weyl phase-space
    variables $x_j^{ab}$, which can be collected into a local phase-space matrix
    $\mathbf{x}_j$.
    (d) The Hubbard-operator algebra induces a Lie--Poisson dynamics generated
    by the Weyl symbol $H_W$, equivalently expressed through a self-consistent
    local effective matrix $\mathbf{h}_j$.
    (e) In $p$TWA, quantum fluctuations are incorporated by sampling an ensemble
    of initial phase-space configurations, evolving each realization
    classically, and averaging the resulting observables over trajectories.
    }
    \label{fig:pTWA_overview}
\end{figure*}

To overcome this difficulty, we employ the local Hubbard operator basis
(shown in Fig.~\ref{fig:pTWA_overview})
\begin{equation}
X_j^{ab}=|a\rangle\langle b|,
\qquad a,b=0,\dots,n-1,
\end{equation}
which forms a complete basis of the local operator space and closes under
commutation,
\begin{equation}
[X_j^{ab},X_j^{cd}]
= \delta_{bc}\,X_j^{ad}-\delta_{ad}\,X_j^{cb}.
\label{eq:Hubbard_algebra}
\end{equation}
The identity component is given by
$\sum_a X_j^{aa}=\mathbf{1}$, while the traceless sector realizes
$\mathfrak{sl}(n)$. The Hubbard operators therefore provide a natural
algebraic structure for a Lie--Poisson phase-space formulation.

At the level of the local multilevel phase space, this construction is closely related to TWA formulations based on Hermitian generalized Gell-Mann operators, in particular the GDTWA of Zhu \emph{et al.}~\cite{Zhu2019}.
The Hubbard basis and the generalized Gell-Mann basis supplemented by the identity are alternative complete bases of the same local operator space and, under the same semiclassical truncation, yield equivalent deterministic equations of motion up to a linear change of coordinates. The Hubbard basis is particularly convenient for Fock parafermions because the local occupation states and occupation-changing operators are represented directly by the matrix units $X_j^{ab}$.

The local $\mathbb{Z}_n$ clock operators $(Z_j, X_j)$ can be expressed as
\begin{equation}
Z_j = \sum_{a=0}^{n-1} \omega^a X_j^{aa},
\qquad
X_j = \sum_{a=0}^{n-1} X_j^{a+1,a}\; (\mathrm{mod}\; n),
\end{equation}
so any Hamiltonian written in terms of parafermions or clock variables can be rewritten entirely in the $X_j^{ab}$ basis. 

We define the classical phase-space variables as the Wigner--Weyl
symbols of the Hubbard operators,
\begin{equation}
x_j^{ab} \equiv \langle X_j^{ab} \rangle_W.
\end{equation}
For $a\neq b$, these variables are generally complex because the
off-diagonal Hubbard operators are non-Hermitian. In units where
$\hbar=1$, the Hubbard-operator commutator gives the Lie--Poisson bracket
\begin{equation}
\{x_j^{ab},x_k^{cd}\}_{\mathrm{PB}}
= -i\,\delta_{jk}
  \left(\delta_{bc}\,x_j^{ad}-\delta_{ad}\,x_j^{cb}\right).
\label{eq:hubbard_poisson}
\end{equation}
Phase-space variables on different sites Poisson-commute, and the
on-site brackets close on the Hubbard basis.

Let $H_W$ denote the Wigner--Weyl symbol of the Hamiltonian. Applying
the chain rule and using Eq.~\eqref{eq:hubbard_poisson}, we obtain
\begin{align}
\dot{x}_j^{ef}
&= \{x_j^{ef},H_W\}_{\mathrm{PB}} \notag\\
&= \sum_{k,a,b}
   \frac{\partial H_W}{\partial x_k^{ab}}
   \{x_j^{ef},x_k^{ab}\}_{\mathrm{PB}} \notag\\
&= -i\sum_{a,b}
   \frac{\partial H_W}{\partial x_j^{ab}}
   \left(\delta_{fa}\,x_j^{eb}-\delta_{eb}\,x_j^{af}\right).
\label{eq:general_LP}
\end{align}

\subsubsection{Gaussian Wigner-function sampling}
Quantum fluctuations enter through stochastic sampling of initial conditions from the Wigner function of the chosen quantum state. For many-body systems with $\mathbb{Z}_n$ parafermions, the exact discrete Wigner function can be cumbersome; a practical alternative is to approximate it by a multivariate Gaussian that reproduces the first and second quantum moments. For each site $j$, define
\begin{equation}
\mu_j^{ab} = \langle X_j^{ab} \rangle, \qquad
C_{jk}^{ab,cd} = \frac{1}{2}\langle \{ X_j^{ab}, X_k^{cd} \} \rangle 
- \langle X_j^{ab} \rangle \langle X_k^{cd} \rangle,
\end{equation}
where $\{\cdot,\cdot\}$ denotes the anticommutator.

The Wigner function is then approximated as a multivariate Gaussian over the phase-space coordinates $x_j^{ab}$, with mean $\mu_j^{ab}$ and covariance $C_{jk}^{ab,cd}$. For the off-diagonal sector, Hermiticity of the Hubbard operators,
$(X_j^{ab})^\dagger=X_j^{ba}$, implies the corresponding constraint
\begin{equation}
x_j^{ba}=(x_j^{ab})^*.
\label{eq:hermiticity_xab}
\end{equation}

For the computational-basis product states used in the simulations below,
the local state at site $j$ is $|a_j\rangle$. In this case the sampling rule
takes a particularly simple closed form:
\begin{equation}
x_j^{aa}(0)=\delta_{a,a_j},
\end{equation}
while, for each $a<b$,
\begin{align}
x_j^{ab}(0)&=q_j^{ab}+i p_j^{ab},\\
x_j^{ba}(0)&=q_j^{ab}-i p_j^{ab},
\end{align}
where
\begin{equation}
q_j^{ab},p_j^{ab}
\sim
\mathcal{N}\!\left(
0,\frac{\delta_{a,a_j}+\delta_{b,a_j}}{4}
\right).
\label{eq:gaussian_sampling_closed}
\end{equation}
Thus, only off-diagonal pairs containing the initially occupied local level
carry fluctuations, while all other off-diagonal components vanish.

Initial classical trajectories $x_{j,\alpha}^{ab}(0)$ are sampled from this Gaussian distribution, ensuring that the quantum first and second moments are reproduced exactly at $t=0$. The subsequent dynamics evolve each trajectory according to the Lie–Poisson equations, and ensemble averaging yields the semiclassical estimate
\begin{equation}
\langle O(t) \rangle \approx 
\frac{1}{N_{\rm traj}} \sum_{\alpha=1}^{N_{\rm traj}} 
O_W\big(\{x_{j,\alpha}^{ab}(t)\}\big).
\end{equation}
This matched-moment Gaussian sampling is analogous to sampling strategies used in other TWA formulations and provides a computationally tractable way of incorporating initial quantum fluctuations. 

\subsubsection{Discrete Wigner-function sampling for $\mathbb{Z}_n$ parafermions}
As an alternative to the Gaussian moment-matching prescription, we also consider a discrete Wigner representation of the finite-dimensional local Hilbert space. In this approach, the initial quantum state is represented by a quasiprobability distribution over a finite set of phase-space points, and expectation values are obtained by averaging the corresponding Weyl symbols. When the discrete Wigner function is non-negative, these phase-space points can be sampled directly as classical initial conditions. This construction is related to, but distinct from, the discrete initial-state sampling employed in
multilevel GDTWA~\cite{Zhu2019}.

A discrete Wigner representation for finite-dimensional Hilbert spaces was developed in several stages: early constructions and systematic classification were reviewed comprehensively by Vourdas~\cite{Vourdas2004}, while Wootters~\cite{Wootters1987} introduced the first fully symmetric discrete Wigner function for prime dimensions, later generalized using finite fields by Gibbons \emph{et al.}~\cite{Gibbons2004} and placed on a group-theoretic foundation by Gross~\cite{Gross2006}. A crucial requirement of these constructions is the existence of a finite field  $\mathbb{F}_n$, which is possible only when $n = p^m$ with $p$ prime. A discrete Wigner function satisfying the standard Wootters axioms (e.g., an $n\times n$ phase space with finite-field structure, translational
covariance, and a striation--MUB correspondence) exists most naturally when the local Hilbert-space dimension is a prime or prime power, $n=p^m$. For composite dimensions that are not prime powers, alternative quasi-probability representations can still be constructed, but they do not
simultaneously retain all of these properties.

\paragraph*{Phase-space structure for prime $n$.}
For a single $\mathbb{Z}_n$ parafermionic site, the local Hilbert space 
$\{|0\rangle,\dots,|n-1\rangle\}$ is $n$-dimensional, and the discrete phase space is the
$n\times n$ grid
$
(q,p)\in\mathbb{F}_n\times\mathbb{F}_n.
$
To each phase-space point one associates a Hermitian \emph{phase-point operator} $A_{qp}$
obeying
\begin{align}
\mathrm{Tr}(A_{qp}) &= 1, \qquad
\mathrm{Tr}(A_{qp}A_{q'p'}) = n\,\delta_{qq'}\delta_{pp'},\\[2pt]
\sum_{q,p\in\mathbb{F}_n} A_{qp} &= n\,\mathbb{I}.
\end{align}
For odd prime $n$, the unique Clifford-covariant Wigner representation~\cite{Gross2006}
expresses $A_{qp}$ in terms of the generalized clock and shift operators
$
Z=\sum_{a=0}^{n-1}\omega^{a}|a\rangle\langle a|,
\qquad
X=\sum_{a=0}^{n-1}|a+1\rangle\langle a|,
\quad
\omega=e^{2\pi i/n},
$
as
\begin{equation}
A_{qp}
=
\frac{1}{n}
\sum_{m,k=0}^{n-1}
\omega^{pk - qm + \tfrac{1}{2} mk}\, Z^{m} X^{k}.
\label{eq:para_Aqp}
\end{equation}
Here $\tfrac{1}{2}$ denotes the multiplicative inverse of $2$ modulo $n$, which exists only for
odd primes (Up to a choice of ordering convention for the Heisenberg--Weyl operators $Z^mX^k$ versus $X^kZ^m$, which only changes the overall phase convention of $A_{qp}$).  The fermionic case $n=2$ requires a different, but closely related, definition~\cite{Wootters1987}.

\paragraph*{Discrete Wigner function.}
Given a density matrix $\rho$ on $\mathcal{H}_n$, its discrete Wigner function is
\begin{equation}
W_\rho(q,p) = \frac{1}{n}\,\mathrm{Tr}\!\big[\rho\,A_{qp}\big],
\end{equation}
a real quasi-probability distribution satisfying $\sum_{q,p}W_\rho(q,p)=1$.  
Wigner negativity signals nonclassicality but is absent for many product states in the
computational basis, allowing $W_\rho$ to serve as a classical probability distribution for
sampling.

\paragraph*{Connection to parafermionic phase-space variables.}
The $p$TWA uses the Hubbard operators $X^{ab}=|a\rangle\langle b|$ as classical phase-space coordinates. For a sampled phase-space point $(q,p)$, the initial classical value is obtained from the
Weyl–Wigner correspondence,
\begin{equation}
x^{ab}(0)
=
\mathrm{Tr}\!\left[A_{qp}\, X^{ab}\right],
\label{eq:xab_from_Aqp}
\end{equation}
which provides the exact Wigner symbol of $X^{ab}$. Since the phase-point operators are Hermitian,
$A_{qp}^\dagger=A_{qp}$, this construction automatically preserves the
Hermiticity constraint, $x^{ba}(0)=[x^{ab}(0)]^*$, for every sampled phase-space point.

\paragraph*{Factorized many-body states.}
If the initial state factorizes across sites,
$\rho=\bigotimes_{j=1}^L \rho_j$, 
then so does its Wigner function:
$ W(q_1,p_1,\dots,q_L,p_L)=
\prod_{j=1}^L W_{\rho_j}(q_j,p_j)$.
One independently samples $(q_j,p_j)$ from each $W_{\rho_j}$ (whenever non-negative) and
converts them into Hubbard coordinates via Eq.~\eqref{eq:xab_from_Aqp}, thereby reproducing
\emph{all} operator moments of the factorized initial state.






\paragraph*{Special case $\boldsymbol{n=3}$.}
For the physically important case of $\mathbb{Z}_3$ parafermions,
the finite-field representation~\eqref{eq:para_Aqp} is equivalent to an SU(3) Wigner construction due to Klimov and de Guise~\cite{Klimov2017}. In this formulation, the phase-point operators can be expressed in terms of a complete set of four mutually unbiased bases (MUBs) in dimension 
$3$ (compatible with the SU(3) structure), providing a natural geometric interpretation compatible with the Hubbard operator basis used in $p$TWA. We employ this SU(3) construction for all $\mathbb{Z}_3$ simulations; details are given in Appendix~\ref{app:SU3Wigner}.

\paragraph*{Computational complexity.}
For a system of $N$ sites with local Hilbert-space dimension $n$, the $p$TWA evolves an $n\times n$ phase-space matrix at each site, corresponding
to $\mathcal{O}(Nn^2)$ dynamical variables and memory cost. For local interactions, the computational cost of evaluating the equations of motion is linear in $N$ at fixed $n$, while the dense local matrix operations have
a generic $\mathcal{O}(n^3)$ cost per site. Long-range interactions may introduce an additional dependence on $N$ according to the interaction
connectivity. 

\begin{figure}[t]
\centering
\includegraphics[width=\linewidth]{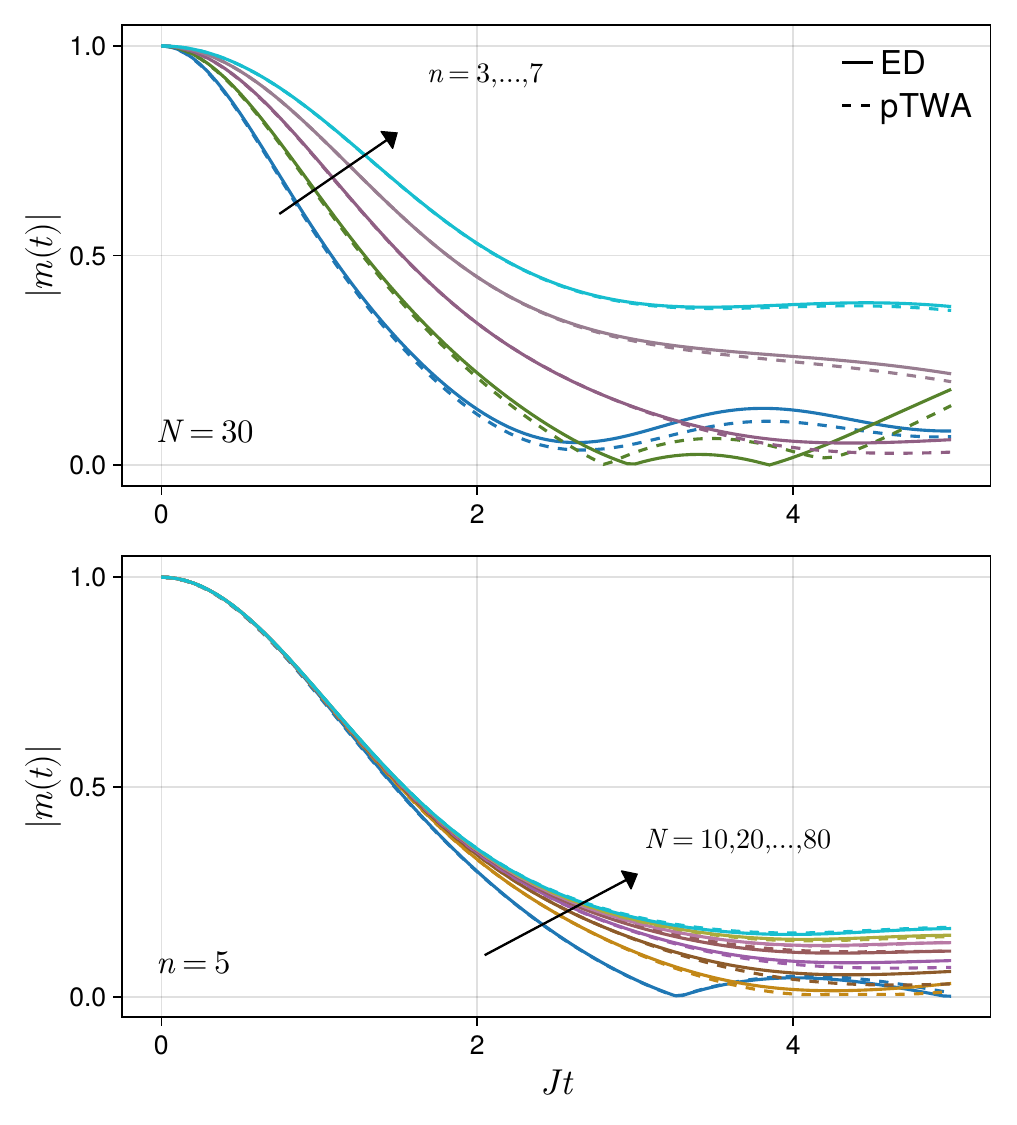}
\caption{Benchmark of the $p$TWA dynamics for the fully connected
$\mathbb{Z}_n$ clock model [Eq.~(\ref{eq:Zn_LMG_H})]. We show the time evolution of the collective magnetization
$|m(t)| = \left| \frac{1}{N}\sum_{i=1}^{N}\langle Z_i \rangle \right|$,
comparing exact diagonalization (solid lines) with $p$TWA
results (dashed lines). The system is initialized in the fully
polarized clock state $|0\rangle^{\otimes N}$.
Model parameters are $J=1.0$ and $g/J=0.5$. Top panel: convergence with increasing local clock dimension
$n=3,4,\dots,7$ at fixed system size $N=30$. Bottom panel: convergence with increasing system size
$N=10,20,\dots,80$ at fixed clock dimension $n=5$. In both cases the agreement between $p$TWA and exact dynamics
improves systematically as the system approaches the semiclassical
limit, either by increasing the local Hilbert-space dimension $n$
or by increasing the number of sites $N$.
}
\label{fig:Zn_LMG_benchmark}
\end{figure}

\section{Benchmark models for the $p$TWA}

Before applying the construction to complex parafermionic models, we first benchmark the underlying multilevel semiclassical formulation in settings where exact solutions are available. These tests are intended to isolate the performance of the local Hubbard-operator Lie--Poisson dynamics and the
initial-state sampling from the additional complications associated with parafermionic exchange statistics. They therefore provide a controlled
validation of the multilevel component of the method that is subsequently used as the local building block of $p$TWA.

\label{sec:III}
\subsection{Single-site $\mathbb{Z}_n$ clock dynamics}
\label{subsec:singleSiteZn}
We first consider the dynamics of a single $\mathbb{Z}_n$ clock degree of freedom driven by a transverse field. The local Hilbert space of a $\mathbb{Z}_n$ clock variable is spanned by
\begin{equation}
|a\rangle , \qquad a=0,1,\dots,n-1 .
\end{equation}

To isolate the local drive dynamics, we consider the Hamiltonian
\begin{equation}
H = -g\left(X+X^\dagger\right).
\label{eq:clock_single_site_H}
\end{equation}
In the $Z$ basis this operator acts as
\begin{equation}
H|a\rangle = -g\left(|a+1\rangle + |a-1\rangle\right),
\end{equation}
so the problem is equivalent to a quantum particle hopping on a periodic ring of $n$ sites. Moreover the dynamics can also be obtained directly from the equation of motion for the density matrix, we present the eigenmode solution because it provides a compact analytic expression and makes the connection to tight-binding dynamics on a periodic ring explicit.

\paragraph{Exact solution.} Because the Hamiltonian is translationally invariant in the clock index,
it is diagonalized by discrete Fourier modes
\begin{equation}
|k_m\rangle =
\frac{1}{\sqrt n}
\sum_{a=0}^{n-1}
e^{ik_m a}|a\rangle ,
\qquad
k_m = \frac{2\pi m}{n},
\end{equation}
where $m=0,\dots,n-1$. Acting with the shift operator yields $X|k_m\rangle = e^{-ik_m}|k_m\rangle$. 
The Hamiltonian~(\ref{eq:clock_single_site_H}) is thus diagonal in this basis,
\begin{equation}
H|k_m\rangle = E_m |k_m\rangle ,
\end{equation}
with eigenenergies $E_m = -2g\cos(k_m)$.
If the system is initialized in a clock basis state $|\psi(0)\rangle = |a_0\rangle$, the transition amplitude to state $|a\rangle$ is
\begin{equation}
A_{a,a_0}(t)
=
\frac{1}{n}
\sum_{m=0}^{n-1}
\exp
\left[
i\,2gt\cos\!\left(k_m\right)
\right]
\exp
\left[
ik_m(a-a_0)
\right].
\label{eq:clock_general_amplitude}
\end{equation}
The exact occupation probabilities are therefore
\begin{equation}
P_{a,a_0}(t)=
\left|
A_{a,a_0}(t)
\right|^2 .
\label{eq:clock_probability}
\end{equation}
Because of translation invariance,
the dynamics depends only on the relative index
$r=a-a_0 \pmod n$. 
Using the identity
\begin{equation}
e^{iz\cos\theta}
=
\sum_{\ell=-\infty}^{\infty}
i^\ell J_\ell(z)e^{i\ell\theta},
\end{equation}
one finds that the transition amplitude becomes $A_r(t)= i^r J_r(2gt)$, leading to $P_r(t)=J_r^2(2gt)$, which corresponds to coherent spreading on an infinite ring.

\paragraph{$p$TWA formulation. } In the local Hubbard basis the shift operator can be written as $X=\sum_{a=0}^{n-1}X^{a+1,a}$, with indices understood modulo $n$. The Hamiltonian then takes the form
\begin{equation}
H=-g\sum_{a=0}^{n-1}
\left(
X^{a+1,a}+X^{a,a+1}
\right).
\end{equation}
Within the $p$TWA framework, we introduce classical phase-space variables $x^{ab}=\langle X^{ab}\rangle_W ,$ which can be collected into an $n\times n$ matrix $x=(x^{ab})$.
The Weyl symbol of the Hamiltonian becomes
\begin{equation}
H_W=-g\sum_{a=0}^{n-1}
\left(
x^{a+1,a}+x^{a,a+1}
\right).
\end{equation}
The classical phase-space variables evolve according to Eq.~\ref{eq:general_LP}. For the single-site Hamiltonian considered here one obtains
\begin{equation}
h_{ba}=-g\left(\delta_{b,a+1}+\delta_{b+1,a}\right),
\end{equation}
so that
\begin{equation}
h=-g(T+T^\dagger),
\end{equation}
with $T_{ba}=\delta_{b,a+1}$.
Because this matrix is time independent,
the classical solution can be written explicitly as
\begin{equation}
x(t)=e^{iht}x(0)e^{-iht}.
\end{equation}
This evolution has the same structure as the unitary evolution of the density matrix, and in fact the phase-space variables correspond to its matrix elements, $x^{ab}(t)=\langle a|\rho(t)|b\rangle$. For an initial clock basis state $|a_0\rangle$ the corresponding
initial phase-space configuration is $x^{ab}(0)=\delta_{a,a_0}\delta_{b,a_0}$,
which corresponds to a projector onto that state.
In contrast to continuous bosonic Wigner representations, where uncertainty relations enforce a finite spread in phase space, the present finite-dimensional phase space associated with the Hubbard-operator algebra allows such a projector to be represented exactly by a single phase-space point.
For this initial basis state the phase-space representation reduces to a single deterministic configuration, so no sampling over initial conditions is required. This absence of sampling is therefore a special property of this initial state and representation, rather than a generic feature of the $p$TWA. The semiclassical populations are therefore obtained from the
diagonal elements $P_a^{\mathrm{pTWA}}(t)=x^{aa}(t)$. With the identification above, this corresponds to the exact occupation probabilities $P_a(t)=\langle a|\rho(t)|a\rangle$, i.e., the modulus squared of the wavefunction amplitudes.

\subsection{Fully connected $\mathbb{Z}_n$ clock model: analogue of the LMG model}
\label{subsec:Zn_LMG}

To further benchmark the $p$TWA formulation, it is useful to consider an interacting model in which semiclassical dynamics becomes asymptotically exact.
A natural choice is the fully connected $\mathbb{Z}_n$ clock model,
which plays the same role for clock variables as the
Lipkin--Meshkov--Glick (LMG) model does for spin systems~\cite{Lipkin1965}.
Like the LMG model, this system possesses permutation symmetry and
reduces to a self-consistent mean-field problem in the thermodynamic
limit.

The model is defined by the Hamiltonian
\begin{equation}
H =
-\frac{J}{N}\sum_{i<j}
\left(
Z_i Z_j^\dagger + Z_i^\dagger Z_j
\right)
-
g\sum_{i}
\left(
X_i + X_i^\dagger
\right),
\label{eq:Zn_LMG_H}
\end{equation}
where $N$ is the number of sites.
The $1/N$ normalization ensures a well-defined thermodynamic limit,
in analogy with the standard LMG Hamiltonian.

It is convenient to introduce the collective clock operator $
M = \sum_{i=1}^{N} Z_i$. Using the identity
\begin{equation}
\sum_{i<j}
\left(
Z_i Z_j^\dagger + Z_i^\dagger Z_j
\right)
=
\left|
\sum_i Z_i
\right|^2 - N ,
\end{equation}
the Hamiltonian~(\ref{eq:Zn_LMG_H}) can be written,
up to an additive constant, as
\begin{equation}
H =
-\frac{J}{N} |M|^2
-
g\sum_i (X_i + X_i^\dagger).
\label{eq:Zn_LMG_collective}
\end{equation}

This form highlights the close correspondence with the LMG model,
whose Hamiltonian can be written as
\begin{equation}
H_{\mathrm{LMG}}
=
-\frac{J}{N} S_x^2
-
h S_z ,
\end{equation}
with $S_\alpha = \sum_i \sigma_i^\alpha$.
In both cases the interaction term depends only on a collective
operator ($S_x$ for spins, $M$ for clock variables), which implies
permutation symmetry and collective dynamics.

In the large-$N$ limit the dynamics becomes mean-field exact.
Introducing the order parameter
$m = \frac{1}{N} \sum_i \langle Z_i \rangle$, the interaction term generates an effective single-site Hamiltonian
\begin{equation}
H_{\mathrm{MF}}
=
-J\left(
m Z^\dagger + m^* Z
\right)
-
g (X + X^\dagger).
\label{eq:Zn_MF}
\end{equation}
Thus the many-body dynamics reduces to a self-consistent evolution
of a single clock degree of freedom in the mean field produced by
the collective order parameter $m$.

Within the $p$TWA framework we again introduce classical phase-space
variables $x_i^{ab} = \langle X_i^{ab} \rangle_W ,$ with $X_i^{ab}=|a\rangle_i\langle b|$.
The Weyl symbols of the clock operators are
\begin{align}
z_i = \sum_{a=0}^{n-1} \omega^a x_i^{aa}, \quad
\chi_i = \sum_{a=0}^{n-1} x_i^{a+1,a}.
\end{align}

The Weyl symbol of the Hamiltonian becomes
\begin{equation}
H_W
=
-\frac{J}{N}
\sum_{i<j}
\left(
z_i z_j^* + z_i^* z_j
\right)
-
g \sum_i
(\chi_i + \chi_i^*).
\end{equation}

Defining the collective field $m=\frac{1}{N}\sum_i z_i,$
and collecting the local phase-space variables into the matrix
\begin{equation}
\mathbf{x}_i =\sum_{a,b}x_i^{ab}|a\rangle\langle b|,\qquad (\mathbf{x}_i)_{ab}=x_i^{ab},
\end{equation}
the classical equations of motion can be written compactly in matrix form as $\dot{\mathbf{x}}_i=i[\mathbf{h}_i,\mathbf{x}_i],$ where the local mean-field matrix $\mathbf{h}_i$ has elements
\begin{equation}
(\mathbf{h}_i)_{ba}= -J\left(m\,\omega^{-a}+m^*\omega^{a}\right)\delta_{ab}-g\left(\delta_{b,a+1}+\delta_{b+1,a}\right).
\end{equation}

\begin{figure}[t]
\centering
\includegraphics[width=\linewidth]{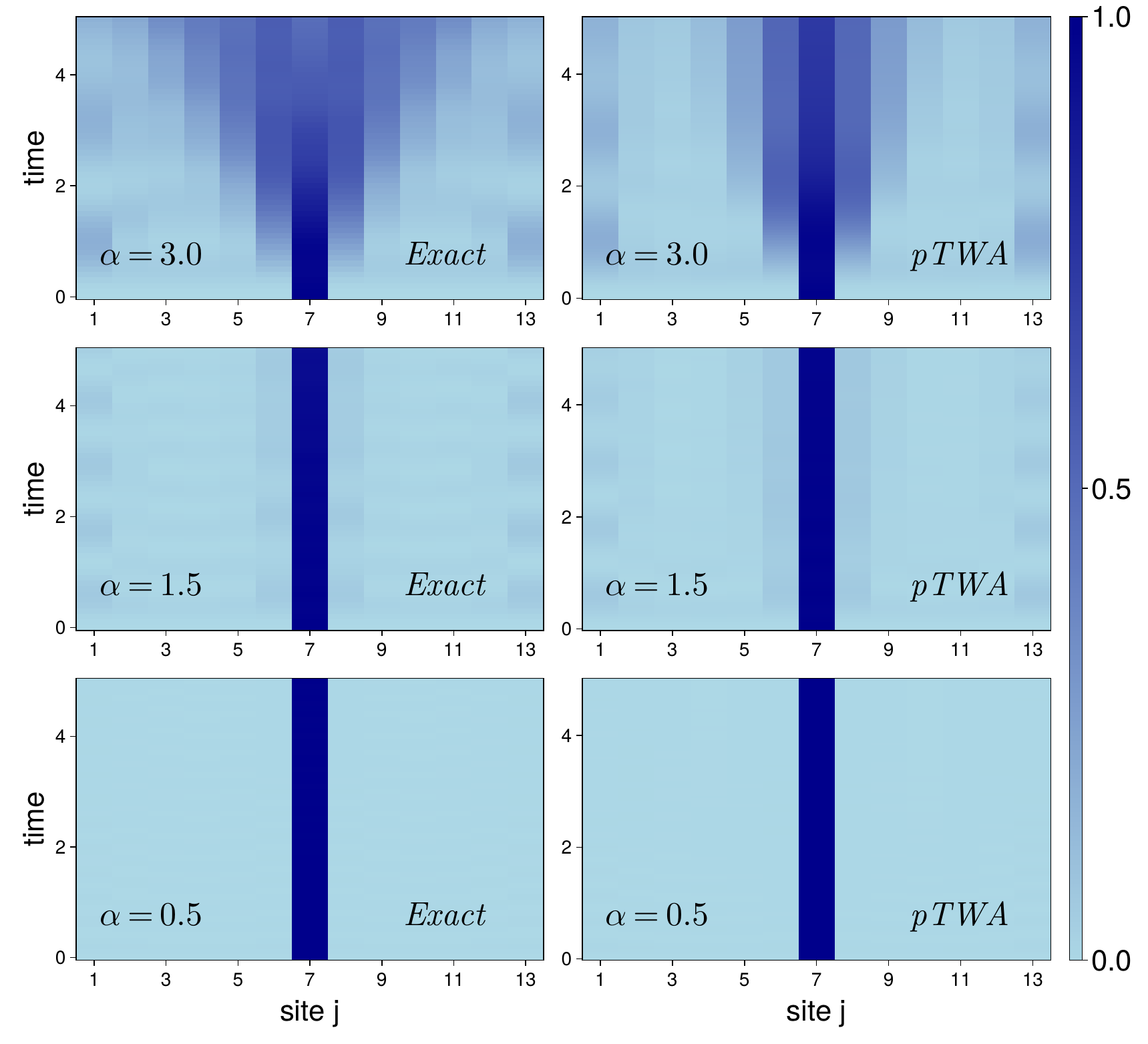}
\caption{Spatiotemporal propagation of a local excitation in the long-range $\mathbb{Z}_3$ clock chain.
We show the excitation probability
$P_{\mathrm{exc}}(j,t)=\langle X_j^{11}(t)+X_j^{22}(t)\rangle$ following an initial excitation prepared at the center site $j_0=7$ in a chain of length $L=13$.
The left column shows exact quantum dynamics obtained from Krylov-based exact time evolution (ED)~\cite{Javad2022}, while the right column shows semiclassical dynamics computed using the parafermionic truncated Wigner approximation ($p$TWA). Rows correspond to different interaction exponents:
$\alpha=3.0$ (top), $\alpha=1.5$ (middle), and $\alpha=0.5$ (bottom). For larger $\alpha$, interactions are effectively short ranged and the excitation spreads outward with a visible light-cone structure. As $\alpha$ decreases and interactions become increasingly long ranged,
the propagation becomes more collective and the excitation distribution
remains strongly concentrated near the initially excited site. We use open boundary conditions.}
\label{fig:lightcone_Z3}
\end{figure}

These equations describe a self-consistent precession of the local
phase-space matrix $\mathbf{x}_i$ in a collective field generated by the order parameter $m$. Because the interaction is infinite range, fluctuations around the mean field vanish as $1/N$, and the semiclassical dynamics becomes exact in the thermodynamic limit. For this reason the fully connected $\mathbb{Z}_n$ clock model provides a natural interacting benchmark for the $p$TWA method.

The model contains two independent parameters controlling the
approach to the semiclassical regime: the system size $N$ and the
local clock dimension $n$.
The large-$N$ limit suppresses collective fluctuations of the order
parameter $m$, rendering the mean-field description asymptotically
exact in the same way as in the LMG model.
Independently, increasing the local dimension $n$ makes the clock
degree of freedom progressively more classical, since the discrete
clock states approach a continuous rotor variable in the limit
$n\to\infty$.

These two limits play complementary roles in validating the $p$TWA
description.
For fixed small $n$, increasing $N$ reduces fluctuations around the
mean-field dynamics and improves the accuracy of semiclassical
approximations.
Conversely, for fixed system size $N$, increasing the local Hilbert
space dimension $n$ provides a controlled route toward a classical
phase-space description.
This property makes the fully connected $\mathbb{Z}_n$ clock model
particularly suitable for benchmarking the $p$TWA method.

In practice we exploit the permutation symmetry of the Hamiltonian
(\ref{eq:Zn_LMG_H}) to perform exact time evolution in the totally
symmetric subspace, which is the natural analogue of the Dicke sector
for the LMG model.
The dimension of this subspace is $\mathrm{dim}_{\mathrm{sym}}(N,n)
=\binom{N+n-1}{n-1},$ which grows polynomially with $N$ for fixed $n$.
This reduction allows exact numerical calculations for system sizes
up to $N\sim 80$ for $n\le7$, providing a useful benchmark against
$p$TWA simulations.

Figure~\ref{fig:Zn_LMG_benchmark} illustrates the accuracy of the
$p$TWA description for the fully connected $\mathbb{Z}_n$ clock model.
We compare the semiclassical dynamics of the collective order parameter
$m_z(t) = \frac{1}{N}\sum_i \langle Z_i \rangle$
with exact diagonalization performed in the permutation-symmetric
subspace.

Two complementary routes toward the semiclassical regime are explored.
In the upper panel we fix the system size ($N=30$) and increase the
local clock dimension $n$. As $n$ increases the discrete clock
variable approaches a continuous rotor, and the $p$TWA dynamics
rapidly converges toward the exact result.

In the lower panel we fix the local dimension ($n=5$) and increase
the system size $N$. Because the interaction is fully connected,
fluctuations of the collective order parameter scale as
$1/\sqrt{N}$, and the semiclassical approximation becomes
asymptotically exact in the thermodynamic limit.
The numerical results clearly show this systematic improvement
with increasing $N$.

Together these results confirm that the $p$TWA formulation
captures the semiclassical dynamics of interacting
$\mathbb{Z}_n$ clock systems.

\section{One-dimensional lattice models}
\label{sec:IV}

\subsection{Long-range $\mathbb{Z}_3$ clock chain}
\label{subsec:LR_clock}

Having validated the $p$TWA formulation in the exactly solvable single-site problem and in the mean-field–controlled fully connected clock model, we now turn to  interacting one-dimensional lattice systems. As a first example we consider a long-range interacting $\mathbb{Z}_3$ clock chain. This model provides a minimal interacting system with a three-dimensional local Hilbert space and nontrivial quantum transport dynamics, while remaining simple enough to permit exact numerical time evolution for moderate system sizes.


We consider the long-range Hamiltonian
\begin{equation}
H
=
-\sum_{i<j} J_{ij}
\big(
Z_i Z_j^\dagger + Z_i^\dagger Z_j
\big)
-
g_x \sum_j (X_j + X_j^\dagger),
\label{eq:clock_H}
\end{equation}
where the couplings decay algebraically as
$J_{ij} = J/|r_i-r_j|^{\alpha}$.
Throughout this work we measure energies in units of $J$ ($J\equiv1$).
The exponent $\alpha$ interpolates between short-range interactions
($\alpha\to\infty$) and strongly long-range regimes. The first term describes ferromagnetic interactions in the clock basis, favoring alignment of the complex clock variables $Z_j$. The transverse-field term (fixed throughout this work to $g_x/J=0.5$)
generates cyclic transitions $|0\rangle \leftrightarrow |1\rangle \leftrightarrow |2\rangle$ ,
thereby driving coherent quantum dynamics and enabling the propagation of local excitations along the chain.



The Weyl symbol of the Hamiltonian becomes
\begin{equation}
H_W
=
-\sum_{i<j} J_{ij}
\big(
z_i z_j^* + z_i^* z_j
\big)
-
g_x \sum_j
\big(
\chi_j + \chi_j^*
\big),
\label{eq:clock_HW}
\end{equation}
where
\begin{align}
z_j &= \sum_{a=0}^{2} \omega^a x_j^{aa}, \\
\chi_j &= x_j^{10} + x_j^{21} + x_j^{02}.
\end{align}

The classical phase-space variables evolve according to the
Lie--Poisson equations inherited from the Hubbard operator algebra,
\begin{equation}
\dot{x}_j^{ef}
= i\sum_{ab}
\frac{\partial H_W}{\partial x_j^{ab}}
\big(\delta_{fa} x_j^{eb}-\delta_{eb} x_j^{af}\big).
\end{equation}

Collecting the local phase-space variables into the matrix
$\mathbf{x}_j$, the equations of motion take the compact form
\begin{equation}
\dot{\mathbf{x}}_j = i[\mathbf{h}_j,\mathbf{x}_j],
\end{equation}
where $\mathbf{h}_j$ is the local effective matrix defined by
\begin{equation}
(\mathbf{h}_j)_{ba}
= \frac{\partial H_W}{\partial x_j^{ab}}.
\label{eq:local_effective_matrix}
\end{equation}
Here $H_W$ is the scalar Weyl symbol of the many-body Hamiltonian,
whereas $\mathbf{h}_j$ is a matrix in the local $n$-dimensional space
whose elements depend on the phase-space variables. Thus,
$\mathbf{h}_j$ should be understood as a self-consistent local
effective matrix rather than as an independent Hamiltonian operator.

For the Hamiltonian~\eqref{eq:clock_HW}, the interaction term contributes
only diagonal elements to $\mathbf{h}_j$,
\begin{equation}
(\mathbf{h}_j)_{aa}
=-2 \sum_{k\ne j} J_{jk}
\,\mathrm{Re}(\omega^a z_k^*),
\end{equation}
while the transverse field generates fixed cyclic off-diagonal entries
proportional to $g_x$.

The resulting classical dynamics can therefore be interpreted as a
self-consistent precession of the local $3\times3$ matrices $x_j$
in mean fields generated by the surrounding sites.

As a first dynamical benchmark we study the propagation of a localized excitation. The system is initialized in a product state in which a single site
$j_0$ at the center of the chain is prepared in the excited state $\ket{1}$, while all other sites remain in the vacuum state $\ket{0}$. To characterize the spreading of the excitation we monitor the
excitation density
\begin{equation}
P_{\mathrm{exc}}(j,t)
=
P_1(j,t) + P_2(j,t)
=
\big\langle X_j^{11}(t) + X_j^{22}(t) \big\rangle ,
\end{equation}
which measures the probability that site $j$ is in an excited state ($|1\rangle$ or $|2\rangle$) relative to the vacuum state $|0\rangle$. This quantity is particularly natural for the $\mathbb{Z}_3$ clock model because the transverse field term mixes the excited states $|1\rangle$ and $|2\rangle$.

Within the $p$TWA framework the same observable is obtained from the classical phase-space variables as
$P_{\mathrm{exc}}(j,t) \approx x_j^{11}(t)+x_j^{22}(t)$, and ensemble averaging over stochastic trajectories yields the semiclassical estimate of the quantum expectation value.

Exact dynamics are computed using Krylov-based time evolution in the full Hilbert space of dimension $3^L$. The Hamiltonian is applied matrix-free in the computational basis, exploiting the diagonal structure of the interaction term and the local cyclic action of the $X_j$ operators.
This implementation enables simulations of chains up to $L\approx15$ sites, providing a controlled benchmark for the semiclassical $p$TWA dynamics.

\begin{figure}[t]
\centering
\includegraphics[width=\linewidth]{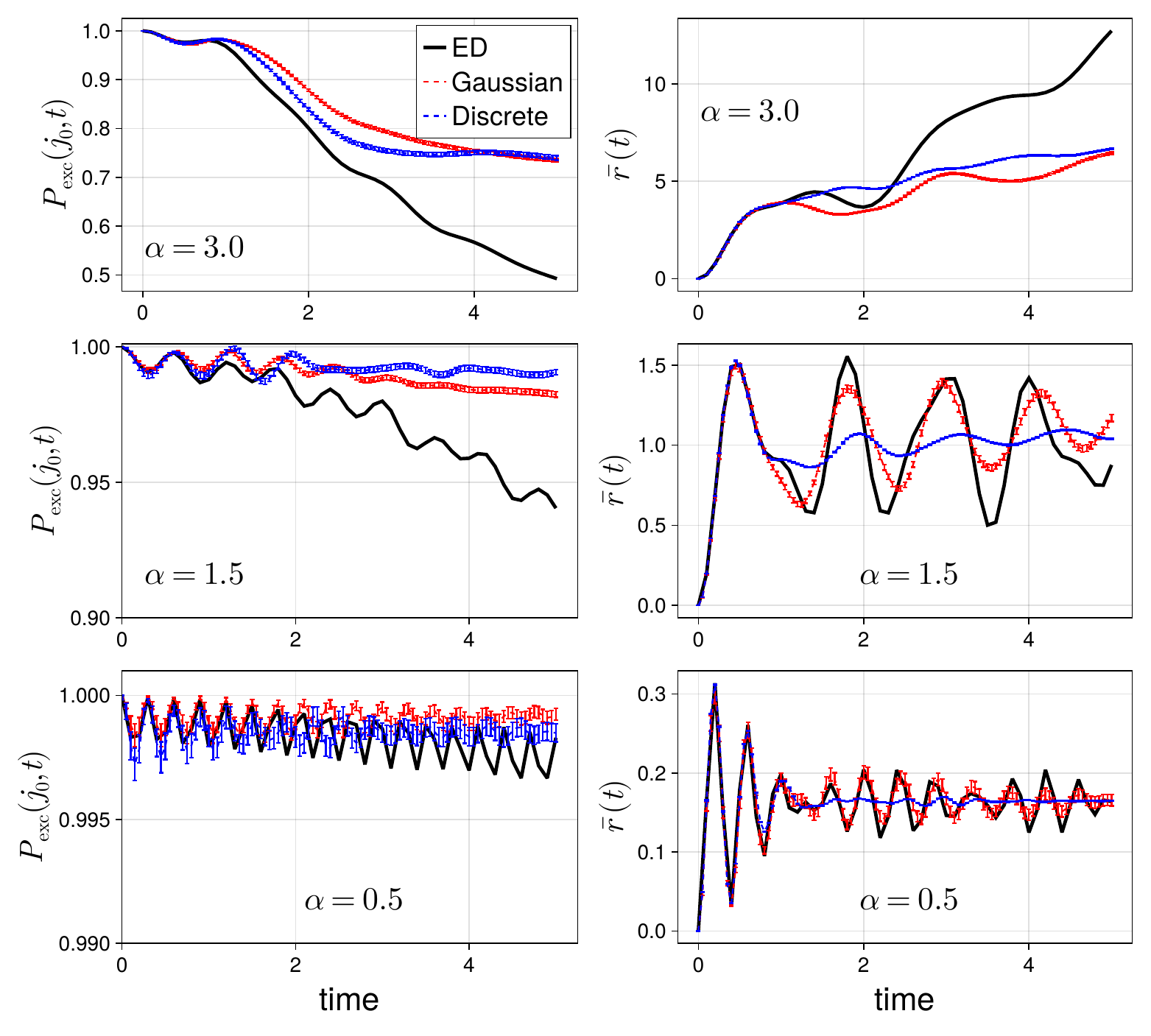}
\caption{ Time evolution of dynamical observables following a local excitation in the long-range $\mathbb{Z}_3$ clock chain of length $L=13$. The system is initialized with a single excitation at the central site $j_0=7$. The left column shows the survival probability $P_{\mathrm{exc}}(j_0,t)$ at the initially excited site, while the right column shows the mean displacement $\bar{r}(t)$, which quantifies the spatial spreading of the excitation. Rows correspond to different interaction exponents: $\alpha=3.0$ (top), $\alpha=1.5$ (middle), and $\alpha=0.5$ (bottom). Black curves denote exact dynamics obtained from exact diagonalization (ED), while red and blue curves show results from the parafermionic truncated Wigner approximation ($p$TWA) using Gaussian and discrete Wigner-function sampling, respectively.}
\label{fig:dynamics_Z3}
\end{figure}

The spatiotemporal structure of the excitation dynamics is illustrated
in Fig.~\ref{fig:lightcone_Z3}, which shows the evolution of the
excitation probability $P_{\mathrm{exc}}(j,t)$ following a local
excitation at the center of the chain.

For relatively short-range interactions ($\alpha=3.0$) the excitation
spreads outward from the initial site with a light-cone-like structure
characteristic of finite-velocity propagation.
As the interaction exponent decreases and the interactions become
more long ranged, the spreading pattern changes qualitatively.
For $\alpha=1.5$ the propagation becomes more diffuse and the front
is no longer sharply defined, while for $\alpha=0.5$ the excitation
remains strongly concentrated near the initially excited site over
the time window considered. Long-range interactions generate weak early-time tails outside the main propagation front. The oscillatory features observed close to the edges of the chain  originate from boundary-induced excitation and are well reproduced by pTWA.

Across all regimes the $p$TWA captures the overall spatiotemporal
features of the exact evolution, indicating that the semiclassical
phase-space description reproduces the main qualitative aspects of
transport in this system.

To quantify these observations we examine two scalar observables derived from the excitation distribution. Figure~\ref{fig:dynamics_Z3} shows the survival probability at the initial site ($j_0=7$), $P_{\mathrm{exc}}(j_0,t)$, together with the mean displacement $\bar{r}(t)=\sum_j |j-j_0|,P_{\mathrm{exc}}(j,t)$, which measures the spatial spreading of the excitation. For $\alpha=3.0$ the excitation decays steadily from the initial site and the mean displacement grows rapidly, reflecting efficient transport across the chain. The semiclassical dynamics reproduces the early-time decay and initial growth but underestimates the long-time spreading observed in the exact dynamics. At intermediate range $\alpha=1.5$ both observables exhibit weak oscillatory features. While long-range couplings enhance the spatial spread of excitations, the oscillations themselves originate from boundary-induced dynamics in the open chain, and are captured qualitatively by the $p$TWA trajectories. For strongly long-range interactions $\alpha=0.5$ the excitation spreads only weakly and the semiclassical results agree closely with the exact evolution over the entire simulated time window.
The relative performance of discrete and Gaussian Wigner sampling depends somewhat on the observable. Discrete sampling tends to reproduce short-time dynamics and oscillatory features slightly more faithfully, while Gaussian sampling provides a comparable description of integrated spreading observables such as the mean displacement.

Overall, these results demonstrate that the $p$TWA framework provides a reliable description of the early- and intermediate-time transport dynamics in long-range $\mathbb{Z}_3$ clock chains. The remaining discrepancies at long times arise primarily from the
mean-field character of the semiclassical approximation rather than
from the choice of Wigner sampling scheme.

\subsection{Disordered $\mathbb{Z}_3$ Fock parafermion chain}
\label{subsec:MBL_parafermion}

Having established the performance of $p$TWA for clean long-range transport in the $\mathbb{Z}_3$ clock chain, we now consider a complementary model in the presence of quenched disorder. Disordered systems provide a qualitatively different dynamical regime: rather than ballistic or diffusive transport, disorder can strongly suppress propagation and lead to long-time memory of the initial state. Testing the semiclassical framework in this regime is therefore important for assessing its ability to capture disorder-induced suppression of transport and long-time memory effects in systems with fractionalized statistics.

We consider a disordered $\mathbb{Z}_3$ Fock parafermion chain
described by the Hamiltonian
\begin{equation}
H = -J \sum_j \Big[
(1-g)\, f_j^\dagger f_{j+1}
+ g\, (f_j^\dagger)^2 f_{j+1}^2
+ \mathrm{h.c.}
\Big]
+ \sum_j \mu_j n_j ,
\label{eq:disordered_parafermion}
\end{equation}

\begin{figure}[t]
  \centering
  \includegraphics[width=\linewidth]{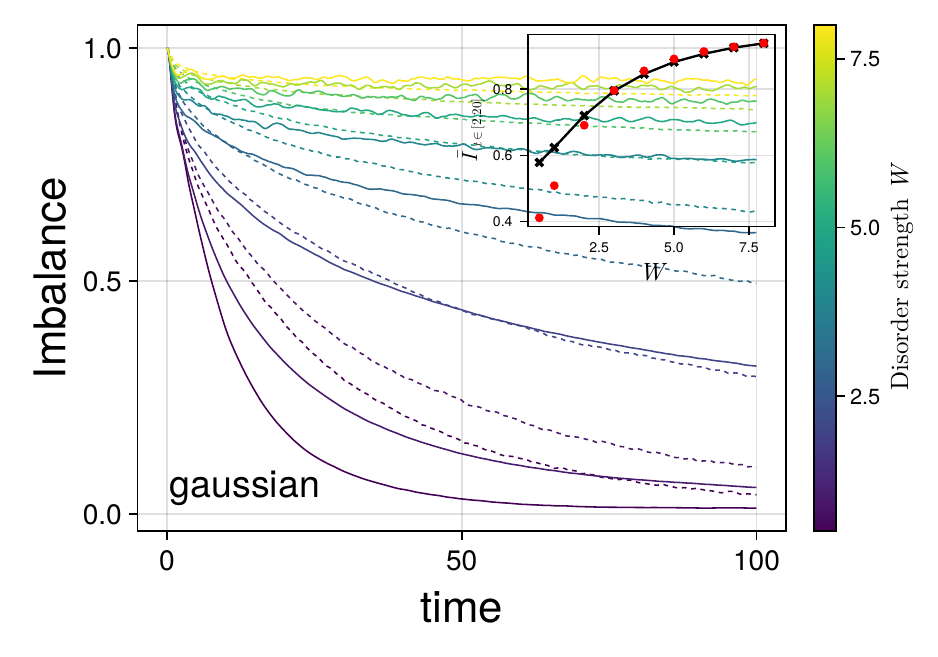}\\
  \includegraphics[width=\linewidth]{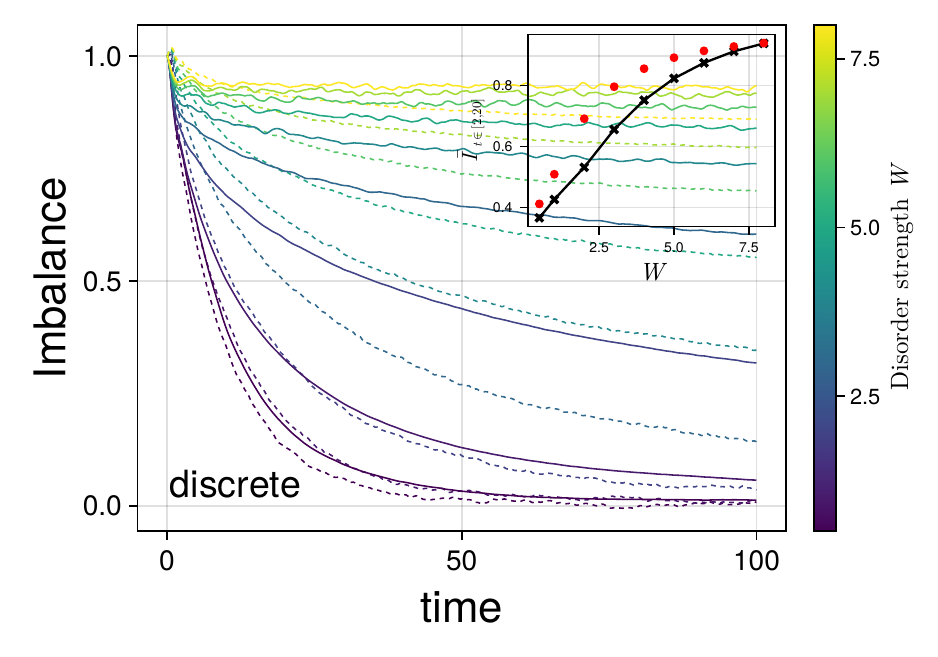}
\caption{
Disorder-averaged imbalance dynamics $\mathcal{I}(t)$ for the disordered $\mathbb{Z}_3$ Fock parafermion chain with system size $L=12$ and pair-hopping parameter $g=0.3$. The top panel shows $p$TWA results obtained using Gaussian Wigner sampling, while the bottom panel uses discrete Wigner sampling. Dashed curves correspond to $p$TWA trajectories and solid curves to exact
diagonalization (ED) benchmarks. Colors indicate the disorder strength $W$. For weak disorder the imbalance decays rapidly toward zero, signaling efficient transport and ergodic relaxation.  Inset: time-averaged imbalance
$\bar{\mathcal{I}}=\langle \mathcal{I}(t)\rangle_{t\in[2,20]}$
as a function of disorder strength $W$. Black $\times$ markers denote $p$TWA results and red circles show
ED benchmarks, revealing a smooth crossover from ergodic transport at weak disorder to strongly suppressed transport at large $W$.}
\label{fig:imbalance_ptwa}
\end{figure}
where the local occupation operator is $n_j = f_j^\dagger f_j + (f_j^\dagger)^2 f_j^2.$ The parameter $g\in[0,1]$ interpolates between single-particle hopping ($g=0$) and coherent pair-hopping processes ($g=1$). The on-site potentials $\mu_j$ represent quenched disorder drawn independently from a uniform distribution $\mu_j\in[-W,W]$.

Although the Hamiltonian~\eqref{eq:disordered_parafermion}
appears quadratic for  $g=0$ in the parafermion operators, the model is intrinsically interacting due to the nontrivial exchange statistics of Fock parafermions, which are implemented through Jordan--Wigner strings. As a consequence, the hopping terms generate effective many-body interactions in the local clock representation. The resulting system is generically nonintegrable and provides a natural platform for studying localization phenomena arising from fractionalized statistics rather than conventional density--density interactions.

Previous numerical studies using exact diagonalization and tensor
network techniques~\cite{Camacho2022,Bahovadinov2022} have shown that
the disordered $\mathbb{Z}_3$ Fock parafermion chain can exhibit
localization-like behavior despite the absence of conventional
density–density interactions.
This makes the model a particularly stringent test for semiclassical
phase-space approaches, which must capture both the suppression of
transport and the persistence of local memory over long times.

To formulate the semiclassical dynamics within the $p$TWA framework, we use
the Fradkin-Kadanoff representation derived in
Appendix~\ref{app:FK_mapping}. For $\mathbb{Z}_3$, the local lowering and
statistical-phase operators are
\begin{equation}
B_j=X_j^{01}+X_j^{12},
\qquad
U_j=X_j^{00}+\omega X_j^{11}+\omega^2X_j^{22},
\end{equation}
and the full Fock-parafermion operator is
\begin{equation}
f_j=
\left(\prod_{\ell<j}U_\ell\right)B_j.
\end{equation}
The nonlocal string implements the fractional exchange statistics.

For the nearest-neighbor terms appearing in
Eq.~\eqref{eq:disordered_parafermion}, the strings cancel pairwise except for
a local statistical factor on site $j$,
\begin{align}
f_j^\dagger f_{j+1}
&=B_j^\dagger U_j B_{j+1},\\
(f_j^\dagger)^2f_{j+1}^2
&=(B_j^\dagger)^2U_j^2B_{j+1}^2.
\end{align}
Thus, although the mapped Hamiltonian is local, the parafermionic exchange
statistics is retained through the factors $U_j$ and $U_j^2$. The local
occupation operator is
\begin{equation}
n_j=X_j^{11}+2X_j^{22}.
\end{equation}
All operators entering the Hamiltonian can therefore be expressed in terms
of the local Hubbard basis and treated using the phase-space construction
introduced in Sec.~\ref{sec:II}.

As an initial condition we consider global quenches from a sharp domain-wall state,
\begin{equation}
|\psi_0\rangle
=
|1,1,\ldots,1,0,0,\ldots,0\rangle ,
\label{eq:domainwall_state}
\end{equation}
in which the left half of the chain is occupied and the right half is empty.
This state provides a sensitive probe of transport: if particles propagate through the system the domain wall melts and the density
profile gradually becomes uniform.
Conversely, persistent spatial imbalance indicates inhibited transport and localization-like dynamics.

To quantify the resulting dynamics we monitor the left and right particle
number imbalance
\begin{equation}
\mathcal{I}(t)=\frac{2}{L}
\Big[N_L(t)-N_R(t)\Big],
\qquad
N_L(t)=\sum_{j=1}^{L/2}\langle n_j(t)\rangle ,
\label{eq:imbalance_def}
\end{equation}
with $N_R(t)$ defined analogously.
By construction $\mathcal{I}(0)=1$.
In ergodic regimes the imbalance decays rapidly toward zero as the
domain wall relaxes, while a finite long-time value signals inhibited
transport and localization-like dynamics.

Within $p$TWA the same observable is evaluated from the classical
phase-space variables through
$n_j(t)\approx x_j^{11}(t)+2x_j^{22}(t)$,
followed by ensemble averaging over stochastic trajectories.
Comparing the time dependence of $\mathcal{I}(t)$ obtained from exact
dynamics and semiclassical simulations therefore provides a direct
test of the ability of $p$TWA to capture localization physics in
systems with fractionalized statistics and quenched disorder.

\begin{figure*}[t]
\centering
\includegraphics[width=\linewidth]{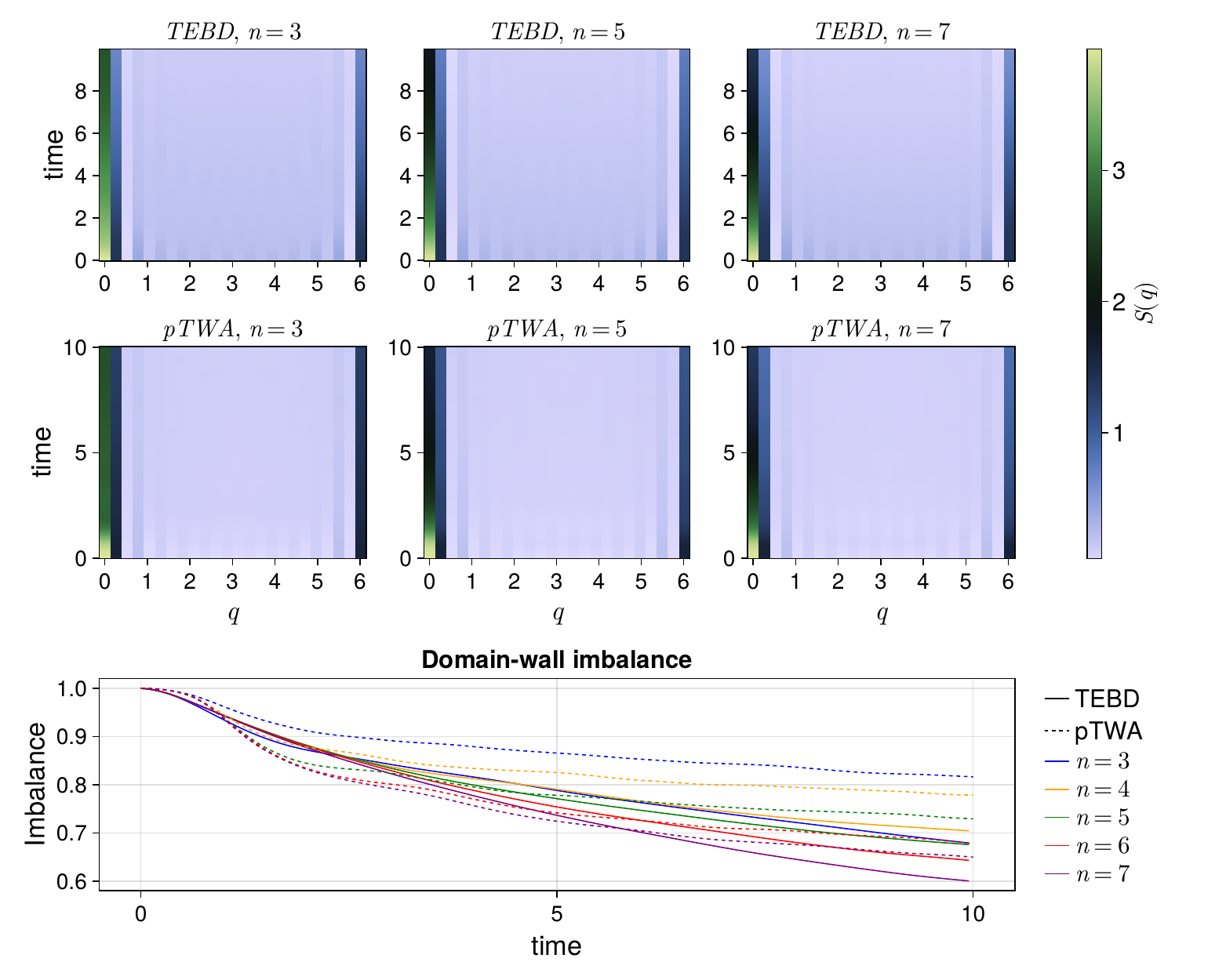}
\caption{Comparison of TEBD and $p$TWA dynamics for the $\mathbb{Z}_n$ chain
($L=24$) initialized in a domain-wall state. Top row: Exact TEBD results for the momentum-resolved Fourier spectrum
$S_0(q,t)$ of the local occupation profile for $n=3,5,7$. Middle row: Corresponding $p$TWA results obtained with Gaussian Wigner sampling using $N_{\mathrm{traj}}=2000$ trajectories. In both approaches the dynamics is generated by the single-hopping Hamiltonian. The heatmaps show the evolution of the momentum distribution of density modulations generated during the relaxation of the domain wall. Bottom panel: Time evolution of the domain-wall imbalance $\mathcal{I}(t)$ for $n=3,\dots,7$.
Solid lines denote TEBD results, while dashed lines correspond to $p$TWA. The semiclassical approach captures the qualitative relaxation dynamics and the dependence on $n$, but overestimates the long-time imbalance due to its mean-field character, while short-time deviations likely arise from Gaussian Wigner sampling, which does not fully capture higher-order correlations of the domain-wall initial state, suggesting that more accurate sampling schemes such as discrete Wigner representations could improve short-time agreement. We verified convergence of the TEBD calculations with respect to both the maximum bond dimension and the Trotter time step.}
\label{fig:Zn}
\end{figure*}

The disorder-averaged imbalance dynamics obtained from $p$TWA is shown in Fig.~\ref{fig:imbalance_ptwa} for a range of disorder strengths $W$. At weak disorder the imbalance decays rapidly toward zero,
reflecting efficient transport and melting of the initial domain-wall profile. As the disorder strength increases the relaxation becomes progressively slower and a pronounced long-time plateau emerges, indicating strong suppression of particle transport and persistent memory of the initial state~\cite{Camacho2022}.

To assess the accuracy of the semiclassical approach we compare the $p$TWA dynamics with exact diagonalization (ED) simulations for identical disorder realizations. Figure~\ref{fig:imbalance_ptwa} shows the disorder-averaged imbalance dynamics obtained using two different sampling schemes within the
$p$TWA framework: Gaussian Wigner sampling (top panel) and discrete Wigner sampling (bottom panel).
In both cases the semiclassical trajectories reproduce the short-time relaxation of the imbalance and capture the systematic increase of the long-time plateau with increasing disorder strength.

The two sampling schemes exhibit small but systematic differences. Discrete Wigner sampling generally shows closer agreement with the
ED dynamics at short times, particularly at weak disorder. This behavior arises because the discrete Wigner representation reproduces the exact operator moments of the product initial state,
whereas Gaussian sampling approximates the initial Wigner function by matching only its first and second moments. Despite this approximation, Gaussian sampling captures the overall disorder dependence and long-time transport suppression across the
entire disorder range.

The inset of Fig.~\ref{fig:imbalance_ptwa} shows the time-averaged
imbalance $\bar{\mathcal{I}}=
\langle \mathcal{I}(t)\rangle_{t\in[2,20]}$, obtained by averaging the imbalance over the time window
$t\in[2,20]$ as a function of disorder strength.
Black $\times$ markers correspond to $p$TWA results, while red circles denote ED benchmarks.
The data reveal a smooth crossover between an ergodic transport regime
at weak disorder and a regime with strongly suppressed transport at large $W$, while showing that the semiclassical dynamics reproduces
the overall disorder dependence of the imbalance.

\subsection{$\mathbb{Z}_n$ Fock Parafermions chain}
\label{subsec:Zn_parafermion}

To extend the analysis beyond the $\mathbb{Z}_3$ case and
to explore larger local Hilbert-space dimensions, we now consider a
$\mathbb{Z}_n$ Fock parafermion chain.
This model allows us to test the $p$TWA framework in a setting where
the local dimension can be varied systematically, providing an
additional route toward the semiclassical limit discussed in the
previous cases. The system is governed by the nearest-neighbor hopping Hamiltonian $H = -J \sum_j \left(
f_j^\dagger f_{j+1} + \mathrm{h.c.}
\right)$, where $f_j$ denotes the $\mathbb{Z}_n$ Fock parafermion operator. Importantly, although this Hamiltonian is bilinear in the Fock-parafermion operators, for $n>2$ it is not a free quadratic hopping problem in local tensor-product degrees of freedom. Under the Fradkin--Kadanoff mapping the hopping term becomes
$f_j^\dagger f_{j+1}=B_j^\dagger U_j B_{j+1},$
where the occupation-dependent operator $U_j$ carries the fractional
statistical phase. Consequently, the model cannot in general be diagonalized by a single-particle Fourier transformation. (see Appendix~\ref{app:FK_mapping} for details).

For the larger local dimensions and system sizes considered here,
exact diagonalization becomes impractical due to the rapid growth of
the Hilbert space ($n^L$).
Instead, we use time-evolving block decimation (TEBD) to obtain
essentially exact quantum dynamics for chains up to $L=24$
sites~\cite{Vidal2003}. The TEBD simulations are performed with open boundary conditions, a time step $\Delta t=0.05/J$, and a maximum bond dimension $\chi_{\max}=80$. We verified convergence for a representative $n=5$ calculation by increasing the bond dimension to $\chi_{\max}=160$ and reducing the time step to $\Delta t=0.025/J$. The corresponding maximum changes in the imbalance over the simulated time interval are $3.2\times10^{-4}$ and $1.0\times10^{-4}$, respectively.

The system is initialized in a domain-wall product state in which
the left half of the chain occupies the local state $\ket{0}$,
while the right half occupies the state $\ket{n-1}$.
This highly inhomogeneous configuration provides a simple probe of
relaxation dynamics and transport across the domain wall.

To characterize the relaxation of the local occupation profile following the domain-wall quench, we analyze the momentum-resolved Fourier spectrum of the local occupation profile. Specifically, in the results shown below we monitor the occupation of the
local state $a=0$, with $P_0=\ket{0}\bra{0},$ 
and compute the momentum-resolved Fourier spectrum
$S_0(q,t)=\frac{1}{L}\left|
\sum_j e^{iqj}\langle P_0(j,t)\rangle\right|^2
$. This quantity measures the spectral weight of density modulations
at wave vector $q$ and therefore provides a convenient diagnostic
of the momentum structure generated during the relaxation process.
In practice, $S_0(q,t)$ is obtained by Fourier transforming the
site-resolved occupation profile $\langle P_0(j,t) \rangle$. In addition, we monitor the domain-wall imbalance $\mathcal{I}(t)=\frac{2}{L}\sum_j s_j\langle P_0(j,t)\rangle$ ,
where $s_j=\pm1$ distinguishes the two halves of the chain. This observable quantifies the decay of the initial density step and
therefore directly measures relaxation across the domain wall.

Figure~\ref{fig:Zn} compares the exact time evolution obtained with
TEBD to $p$TWA simulations using Gaussian Wigner sampling. The upper panels show the momentum-resolved Fourier spectrum
$S_0(q,t)$ of the local occupation profile for $n=3,\dots,7$.
The semiclassical approach reproduces the qualitative momentum
structure and early-time dynamics observed in the exact simulations,
including the redistribution of spectral weight generated by the
domain-wall quench.

The lower panel shows the time evolution of the domain-wall
imbalance $\mathcal{I}(t)$.
The $p$TWA dynamics captures the overall decay of the imbalance
and the qualitative dependence on the clock dimension $n$,
but systematically predicts a slower relaxation,
leading to a larger residual imbalance at long times.

This deviation reflects the mean-field character of the semiclassical
dynamics, which neglects higher-order quantum correlations that
gradually build up during the relaxation process. We observe that the agreement between $p$TWA and the exact dynamics becomes somewhat better for larger values of $n$, although this trend is not uniform across all times. Physically, increasing $n$ enlarges the local Hilbert space and
causes the discrete clock variable to approach a continuous rotor
degree of freedom, thereby enhancing the validity of the
semiclassical phase-space description. This trend is consistent with the semiclassical behavior observed
in the fully connected $\mathbb{Z}_n$ clock benchmark,
where increasing the local dimension $n$ provides a controlled
route toward classical dynamics.

\section{Discussion and outlook}
\label{sec:V}
In this work we introduced the parafermionic truncated Wigner
approximation ($p$TWA), a semiclassical phase-space framework for
simulating the nonequilibrium dynamics of systems with fractionalized
exchange statistics.
The method extends truncated Wigner approaches developed for bosonic
and fermionic systems to $\mathbb{Z}_n$ Fock parafermions by expressing the dynamics in terms of local Hubbard operators, which form a closed Lie algebra on each site.
This representation leads naturally to a Lie--Poisson phase-space
structure and avoids the Grassmann-variable complications that arise
in fermionic Wigner formulations.

Within this framework the quantum dynamics is approximated by
stochastic sampling of initial conditions in the space of classical
phase-space variables $x_j^{ab}=\langle X_j^{ab}\rangle_W$ followed by
deterministic evolution according to the corresponding Lie--Poisson
equations.
We considered both Gaussian Wigner sampling, which reproduces the
first and second moments of the quantum state, and discrete Wigner
sampling based on finite-dimensional phase-space constructions.

The accuracy and range of applicability of the $p$TWA method were
examined through several benchmark problems.
For a single-site $\mathbb{Z}_n$ clock system the semiclassical
equations reproduce the exact quantum dynamics, providing a direct
consistency check of the formalism.
For the fully connected $\mathbb{Z}_n$ clock model, the natural
generalization of the Lipkin-Meshkov-Glick model, the method
captures the collective mean-field dynamics and becomes asymptotically
exact in the semiclassical limit.
We then applied the approach to interacting one-dimensional systems.
For long-range $\mathbb{Z}_3$ clock chains the method reproduces the
qualitative features of excitation spreading and its dependence on
interaction range.
For disordered $\mathbb{Z}_3$ Fock parafermion chains the $p$TWA
dynamics captures the suppression of transport and the emergence of
long-time imbalance plateaus associated with localization-like
behavior.
Finally, simulations of $\mathbb{Z}_n$ Fock parafermion chains show
that the agreement between $p$TWA and exact tensor-network results
improves systematically as the local Hilbert space dimension $n$
increases, approaching the semiclassical limit.

These results demonstrate that the $p$TWA framework provides a useful
and computationally efficient approach for studying the dynamics of
parafermionic systems in regimes where exact numerical methods are
limited by Hilbert-space growth.
At the same time, the method inherits the typical limitations of
semiclassical phase-space approaches: because the dynamics is
effectively mean-field in nature, long-time evolution can deviate
from exact quantum behavior when higher-order correlations become
important.

Several directions for future work naturally follow.
On the methodological side, it would be interesting to develop
improved sampling schemes or higher-order semiclassical corrections
that incorporate quantum fluctuations beyond the truncated Wigner
approximation.
Cluster extensions of the $p$TWA framework, analogous to cluster
truncated Wigner methods previously applied to spin systems~\cite{Wurtz2018,adrian2024},
may provide a systematic way to include short-range correlations and
extend the accuracy of the method to longer times. On the physical side, the phase-space formulation developed here
opens the possibility of studying larger parafermionic systems,
including higher-dimensional clock models, Floquet-driven systems~\cite{Bukov04032015},
and models hosting topological parafermion edge modes\cite{Benhemou2023}.

\section*{Acknowledgment}
J.V.\ acknowledges the Ataro Group for generously providing office space and computational resources that facilitated the completion of this work. M.G.\ is supported by funding from the German Research Foundation (DFG) under the project identifiers 398816777-SFB 1375 (NOA) and 550495627-FOR 5919 (MLCQS), from the Carl-Zeiss-Stiftung within the QPhoton Innovation Project MAGICQ, and from the Federal Ministry of Research, Technology and Space (BMFTR) under project BeRyQC.

\section*{Code availability}
The Julia codes used for the numerical calculations and for reproducing the
figures in this work are publicly available at
\url{https://github.com/javahedi/project-ptwa}.

\appendix
\section{SU(3) Wigner function for $\mathbb{Z}_3$ parafermions}
\label{app:SU3Wigner}

For $\mathbb{Z}_3$ parafermions, the local Hilbert space is 
three-dimensional, $\mathcal{H}_3$, with computational basis
$|0\rangle,|1\rangle,|2\rangle$. Although a discrete Wigner representation can be constructed for any odd prime $n$ using the finite-field approach of 
Refs.~\cite{Wootters1987,Gibbons2004,Gross2006,Vourdas2004},
the $\mathbb{Z}_3$ case admits a particularly natural formulation based on the group SU(3). This construction, developed by Klimov and de~Guise~\cite{Klimov2017},
expresses the Wigner function entirely in terms of projectors onto the four mutually unbiased bases (MUBs) of SU(3), providing a geometrically transparent description especially well suited to the Hubbard-operator phase space used in $p$TWA.

\subsection{Mutually unbiased bases of SU(3)}
Any complete Wigner representation in dimension $d$ requires
$d+1$ mutually unbiased bases when $d$ is prime.
For $d=3$, these are:
$\mathcal{B}^{(b)}=\{\,
|e^{(b)}_k\rangle \colon k=0,1,2 \,\}, 
\qquad b=0,1,2,3,$ satisfying $\bigl|\langle e^{(b)}_k | e^{(b')}_{k'}\rangle\bigr|
=\delta_{bb'}\delta_{kk'}+(1-\delta_{bb'})/\sqrt{3}.$ The first basis is the computational basis,
\begin{equation}
\mathcal{B}^{(0)}:
\qquad 
|e^{(0)}_k\rangle \equiv |k\rangle,\qquad k=0,1,2.
\end{equation}

For $b=1,2,3$, the SU(3) MUB vectors are obtained via quadratic
Gauss sums:
\begin{equation}
|e^{(b)}_k\rangle
=
\frac{1}{\sqrt{3}}
\sum_{n=0}^2 
\omega^{\,b n^2 + k n}\,
|n\rangle,
\qquad
\omega = e^{2\pi i/3}.
\label{eq:SU3_MUB_vectors}
\end{equation}
These four bases define rank-one projectors
\begin{equation}
P^{(b)}_k = |e^{(b)}_k\rangle\langle e^{(b)}_k| ,
\qquad b=0,\dots,3,\quad k=0,1,2.
\end{equation}
Together, $\{P^{(b)}_k\}$ provide a complete operator frame for
$\mathcal{H}_3$.

\subsection{Striations and phase-space geometry}
The discrete phase space consists of pairs 
$(q,p)\in\mathbb{Z}_3\times \mathbb{Z}_3$.
Each point belongs to exactly one line in each of the four \emph{striations}, 
defined by the affine-linear mappings
\begin{align}
\ell_0(q,p) &= q, \\
\ell_1(q,p) &= p, \\
\ell_2(q,p) &= q + p \mod 3,\\
\ell_3(q,p) &= q + 2p \mod 3.
\end{align}
The four striations correspond exactly to the four SU(3) MUBs. Each striation consists of three parallel lines, and each line is assigned one projector from the corresponding MUB.

\subsection{Phase-point operators}

The SU(3) phase-point operators are defined as
\begin{equation}
A_{qp}
=
\frac{1}{3}
\bigg[
\mathds{1}
+ \sum_{b=0}^3 
\Big(
P^{(b)}_{\ell_b(q,p)} 
- \frac{1}{3}\mathbb{I}
\Big)
\bigg],
\label{eq:SU3_Aqp}
\end{equation}
where $(q,p)\in\mathbb{Z}_3^2$.  
This expression satisfies all Wigner axioms:

\paragraph*{(i) Hermiticity}
Each $A_{qp}$ is manifestly Hermitian.

\paragraph*{(ii) Unit trace}
$
\mathrm{Tr}(A_{qp}) = \mathds{1}.
$

\paragraph*{(iii) Orthogonality}
$
\mathrm{Tr}\!\left(A_{qp}A_{q'p'}\right)
=
3 \,\delta_{qq'}\delta_{pp'}.
$

\paragraph*{(iv) Completeness}
$
\sum_{q,p} A_{qp} = 3\,\mathds{1}.
$

\paragraph*{(v) SU(3) covariance}
Under the discrete displacement operators 
$D(m,n)=\omega^{mn/2}X^{m}Z^{n}$ (with $m,n\in\mathbb{Z}_3$),
one has
$
A_{qp}
=
D(q,p)\,A_{00}\,D(q,p)^\dagger.
$
This parallels the Weyl–Heisenberg covariance of the finite-field
construction but is now formulated entirely within SU(3).

\subsection{Wigner function and Hubbard operators}

Given a single-site density matrix $\rho$, the SU(3) Wigner function is
\begin{equation}
W_\rho(q,p)
=
\frac{1}{3}\,
\mathrm{Tr}\!\left[\rho\,A_{qp}\right],
\label{eq:SU3_Wigner}
\end{equation}
a real quasi-probability distribution on the $3\times 3$ phase-space grid.

Since the $p$TWA uses the Hubbard operator basis
$X^{ab}=|a\rangle\langle b|$ as phase-space coordinates, a sampled point
$(q,p)$ is converted to classical initial data via
\begin{equation}
x^{ab}(0)
=
\mathrm{Tr}\!\left[ A_{qp}\,X^{ab} \right].
\label{eq:SU3_Hubbard_projection}
\end{equation}
These values provide the exact Wigner--Weyl symbols of the Hubbard
operators and satisfy all algebraic constraints needed for the SU(3)
Lie--Poisson time evolution.

\subsection{Factorized many-body states}

For an initial many-body state 
$\rho=\bigotimes_{j=1}^{L}\rho_j$, the Wigner function factorizes:
$
W(q_1,p_1,\dots,q_L,p_L)
=
\prod_{j=1}^{L}W_{\rho_j}(q_j,p_j).
$
One samples each $(q_j,p_j)$ independently whenever 
$W_{\rho_j}(q_j,p_j)\ge 0$.
Substituting Eq.~\eqref{eq:SU3_Hubbard_projection} for each site gives
initial conditions that reproduce \emph{all} operator moments of the
factorized state exactly.

\subsection{Relation to finite-field constructions}

For $n=3$, Eq.~\eqref{eq:SU3_Aqp} is equivalent to the
finite-field expression
$
A_{qp}=\frac{1}{3}\sum_{m,k=0}^{2}
\omega^{pk - qm + \tfrac{1}{2}mk}\, Z^{m}X^{k},
$
but the SU(3) MUB formulation is often more natural for parafermionic
systems because:
\begin{enumerate}
\item
it makes explicit the underlying SU(3) structure of the local Hilbert
space;
\item
the projectors $P^{(b)}_k$ correspond directly to extremal points of the
Bloch-body of SU(3);
\item
the mapping to Hubbard operators $X^{ab}$ is particularly simple and
numerically stable.
\end{enumerate}

In this work, all $\mathbb{Z}_3$ simulations employ the SU(3)
Wigner construction of Klimov and de~Guise, 
Eqs.~\eqref{eq:SU3_MUB_vectors}--\eqref{eq:SU3_Aqp}.
This provides an internally consistent, geometrically natural,
and computationally efficient discrete phase-space representation
for parafermionic truncated Wigner dynamics.

\section{Fradkin--Kadanoff representation of the $\mathbb{Z}_n$ Fock-parafermion chain}
\label{app:FK_mapping}

In this Appendix we rewrite the $\mathbb{Z}_n$ Fock-parafermion Hamiltonian
in terms of local operators using the Fradkin--Kadanoff (FK)
transformation. This representation is convenient for numerical
methods such as exact diagonalization and tensor-network algorithms,
as it expresses the model as a chain of finite-dimensional on-site
degrees of freedom with strictly local couplings.
Although the benchmarks in the main text focus on the $\mathbb{Z}_3$
case, the derivation presented below applies generally to
$\mathbb{Z}_n$ parafermion chains.
We consider a $\mathbb{Z}_n$ Fock-parafermion chain described by the Hamiltonian
\begin{equation}
H = -J \sum_j \left[
(1-g) f_j^\dagger f_{j+1}+g(f_j^\dagger)^m f_{j+1}^m+
\mathrm{h.c.}
\right]+
\sum_j \mu_j n_j ,
\label{eq}
\end{equation}
where the local occupation operator isv $n_j = \sum_m (f_j^\dagger)^m f_j^m
$, $g$ controls the
relative weight of single-particle and higher-order hopping processes,
and $\mu_j$ are site-dependent potentials.
The Fock parafermion operators $f_j$ are nonlocal objects from the
point of view of a tensor-product lattice basis, since their exchange
statistics is implemented through nontrivial statistical phases.
For numerical implementations it is therefore advantageous to work with
local operators acting on a finite on-site Hilbert space.
The Fradkin--Kadanoff (FK) transformation, which can be viewed as the
$\mathbb{Z}_n$ generalization of the Jordan--Wigner mapping,
achieves exactly this.
It rewrites parafermion operators in terms of local clock-like
operators acting on an $n$-dimensional on-site Hilbert space.
Each lattice site $j$ carries an $n$-dimensional Hilbert space
spanned by occupation states
\begin{equation}
|n_j\rangle, \qquad n_j = 0,1,\dots,n-1 .
\end{equation}

We introduce two local operators $B_j$ and $U_j$ defined by
\begin{align}
B_j |n_1,\dots,n_j,\dots,n_L\rangle
&=
|n_1,\dots,n_j-1,\dots,n_L\rangle , \\
U_j |n_1,\dots,n_j,\dots,n_L\rangle
&=
\omega^{n_j}
|n_1,\dots,n_j,\dots,n_L\rangle ,
\end{align}
where
\begin{equation}
\omega = e^{2\pi i/n}.
\end{equation}

On the same site these operators satisfy the
$\mathbb{Z}_n$ Weyl algebra
\begin{equation}
U_j B_j = \omega B_j U_j,
\qquad
B_j^n = 0,
\qquad
U_j^n = \mathds{1}.
\label{eq:app_Weyl_algebra}
\end{equation}
Operators acting on different sites commute.

A convenient matrix representation is obtained by taking
\begin{equation}
B =
\begin{pmatrix}
0 & 1 & 0 & \cdots & 0 \\
0 & 0 & 1 & \cdots & 0 \\
\vdots & & & \ddots & \vdots \\
0 & 0 & 0 & \cdots & 1 \\
0 & 0 & 0 & \cdots & 0
\end{pmatrix},
\qquad
U =
\mathrm{diag}(1,\omega,\omega^2,\dots,\omega^{n-1}),
\end{equation}
with $B_j$ and $U_j$ acting as $B$ and $U$ on site $j$ and
as the identity elsewhere.

The Fradkin--Kadanoff mapping expresses the parafermion operators as
\begin{equation}
f_j =
\left(\prod_{\ell=1}^{j-1} U_\ell \right) B_j ,
\qquad
f_j^\dagger =
B_j^\dagger
\left(\prod_{\ell=1}^{j-1} U_\ell^\dagger \right) .
\label{eq:FK_map_general}
\end{equation}
The string $\prod_{\ell<j} U_\ell$ is the $\mathbb{Z}_n$ analogue of
the Jordan--Wigner string and encodes the parafermionic exchange
statistics.

We now rewrite the hopping terms of the Hamiltonian in terms of the
local operators $B_j$ and $U_j$.

\paragraph{Single-particle hopping.}

Using Eq.~\eqref{eq:FK_map_general}, we obtain
\begin{align}
f_j^\dagger f_{j+1}
&=
B_j^\dagger
\left(\prod_{\ell=1}^{j-1} U_\ell^\dagger \right)
\left(\prod_{\ell=1}^{j} U_\ell \right)
B_{j+1} \\
&=
B_j^\dagger U_j B_{j+1},
\end{align}
since all string operators with $\ell<j$ cancel pairwise.
Therefore the hopping term becomes
\begin{equation}
f_j^\dagger f_{j+1} + \mathrm{h.c.}
=
B_j^\dagger U_j B_{j+1}
+
B_{j+1}^\dagger U_j^\dagger B_j .
\end{equation}

\paragraph{Pair hopping.}

A similar calculation gives
\begin{align}
(f_j^\dagger)^m f_{j+1}^m
&=
(B_j^\dagger)^m
\left(\prod_{\ell=1}^{j-1} U_\ell^\dagger \right)^m
\left(\prod_{\ell=1}^{j} U_\ell \right)^m
B_{j+1}^m \\
&=
(B_j^\dagger)^m U_j^m B_{j+1}^m .
\end{align}
Thus the pair-hopping contribution becomes
\begin{equation}
(f_j^\dagger)^m f_{j+1}^m + \mathrm{h.c.}
=
(B_j^\dagger)^m U_j^m B_{j+1}^m
+
(B_{j+1}^\dagger)^m U_j^{\dagger m} B_j^m .
\end{equation}

Collecting the results, the Hamiltonian can be written entirely in
terms of local operators as
\begin{equation}
\begin{aligned}
H = &-J \sum_j
\Big[
(1-g)\, B_j^\dagger U_j B_{j+1}
+ g\, (B_j^\dagger)^m U_j^m B_{j+1}^m
+ \mathrm{h.c.}
\Big] \\
&+ \sum_j \mu_j n_j .
\end{aligned}
\end{equation}

\paragraph{Numerical implementation.}
The local FK representation is used as the starting point for both
exact diagonalization and tensor-network simulations.
For the nonequilibrium dynamics reported in the main text we evolve the
initial product state using a matrix-product-state (MPS)
representation and a standard real-time TEBD scheme with a
second-order Suzuki--Trotter decomposition of the nearest-neighbor FK
Hamiltonian.
At each time step the state is truncated to a maximum bond dimension
$\chi_{\max}$, providing a controlled approximation whose accuracy can
be systematically improved by increasing $\chi_{\max}$ and decreasing
the time step $\delta t$.
In the calculations shown here we used $\delta t=0.05$ and checked that
the observables of interest are converged with respect to both
$\delta t$ and $\chi_{\max}$ within the plotted time window.
For small system sizes we additionally compared the MPS evolution with
exact diagonalization, finding excellent agreement.
For this reason TEBD serves as the numerical benchmark for the larger
chains considered in the main text: unlike $p$TWA, whose error is
intrinsic to the semiclassical approximation, the TEBD error can be
quantified and systematically reduced by standard convergence tests.

\bibliographystyle{apsrev4-2} 
\bibliography{ref/references} 
\end{document}